# Polywell Revisited

Jaeyoung Park[1*], Nicholas A. Krall[1], Giovanni Lapenta[2], and Masayuki Ono[3]

1. Energy Matter Conversion Corporation (EMC2), San Diego, CA, USA
2. Department of Mathematics, KU Leuven, Leuven, Belgium
3. Princeton Plasma Physics Laboratory, Princeton, NJ, USA

*Corresponding Author: Jaeyoung Park, jypark@emc2fusion.com

ORCID: Jaeyoung Park (0000-0002-8360-4568), Giovanni Lapenta (0000-0002-3123-4024), and Masayuki Ono (0000-0001-9849-9417)

**Abstract**

The Polywell fusion concept, originally proposed by Robert W. Bussard in 1985, has been investigated for over four decades as a potential solution for achieving net fusion energy in a compact and economically viable reactor. It combines two distinct approaches: high-beta magnetic cusp confinement of electrons using polyhedral coil configurations and electrostatic ion confinement via a potential well formed by injected electron beams. While the hybrid nature of the Polywell system offers advantages in plasma stability and engineering simplicity, previous efforts have been limited by persistent challenges in achieving sufficient plasma confinement required to generate a net energy gain. In this study, we examine previous work and identify limitations of several Polywell embodiments that have historically impeded progress. We present an updated Polywell physics model incorporating experimental findings and recent first-principles particle-in-cell simulations. This updated model outlines a credible path toward overcoming confinement losses and achieving net energy gain using deuterium-tritium (D-T) fuels. Our findings provide a renewed scientific basis for the continued development of the Polywell fusion concept as a practical and scalable approach to fusion energy.

# 1. Introduction

In 1985, Robert W. Bussard introduced a novel fusion approach for compact and economical fusion power systems, known as the Polywell, a portmanteau of "polyhedral cusp" and "electrostatic potential well." His idea, disclosed in U.S. Patent No. 4,826,646, combines the merits of high beta magnetic cusp confinement and inertial electrostatic confinement (IEC) to achieve sufficiently high plasma confinement in fusion conditions for net energy production [1].

The Polywell concept relies on the following three essential physical mechanisms:

M1: Magnetic cusp confinement of electrons in a high beta plasma.

The first and foremost important part of the Polywell approach is highly efficient plasma confinement in the magnetic cusp, operating in a high-beta plasma, known as the high-beta cusp. It is based on a theoretical conjecture by Grad and colleagues at New York University in the 1950s, which proposed that plasma particles could be specularly reflected, and their confinement greatly enhanced by the sharp boundary layer produced at the surface of a magnetic cusp operating at a high beta [2]. Key figures of merit for the high beta cusp fusion approach are its intrinsic plasma stability and compact reactor size. For example, a high beta cusp reactor can potentially produce multiple gigawatts of D-T fusion power in a 40 cm radius device operating with a 10 T magnetic field and 20 keV plasma temperature for ions and electrons, owing to its ability to sustain plasma densities on the order of $\sim 10^{22}$ ions/m$^3$ [2]. This compact device would achieve breakeven if its confinement favorably scales with the electron gyroradius. This attracted multiple research groups in the 1960s and 1970s to investigate the plasma dynamics in a cusp system [3,4]. However, cusp fusion research was largely discontinued in the early 1980s due to uncertainties about achieving the necessary confinement and technical difficulties in generating high-beta cusp conditions during plasma start-up. In particular, the start-up issue is critical for the high beta approach because plasma losses are significant when beta is small. Unless the input power is sufficiently high during start-up, the lossy nature of the cusp system in the low-beta state prevents reaching the high-beta regime, where Grad and his colleagues predicted enhanced confinement. Note that the Polywell approach requires the high-beta cusp to confine energetic electrons magnetically as a first step in its operation.

M2: Formation of an electrostatic potential well via electron beam injection.

Once energetic electron confinement is established in the high-beta cusp, the injection of electron beams through the magnetic cusp openings creates an excess of electrons in the central region, leading to the formation of an electrostatic potential well. Though not explicitly discussed by Bussard in his patent [1], the confinement efficiency of electron beams is critical in producing and sustaining a potential well on the order of tens of keV for fusion power generation. This is why it is crucial for a Polywell device to achieve a high beta condition during its start-up prior to potential well formation through electron beam injection.

M3: Electrostatic confinement of ions by the potential well and synergistic confinement improvement for magnetic cusp confinement

The final essential mechanism involves the confinement of positively charged fusion fuel ions by the potential well. The potential well confines positive fusion fuel ions by slowing them down at the plasma boundary and returning them to the center at high energy. At the cusp boundary, the ion gyroradius is also significantly reduced due to the potential well. The hypothesis is that the presence of a potential well would enhance ion confinement beyond the conventional high-beta cusp system, which in turn would improve electron confinement due to the ambipolar constraint. This final mechanism completes the hybrid plasma confinement hypothesis of Polywell, magnetic confinement for electrons and electrostatic confinement for ions, provided that a high beta cusp and a power-efficient potential well sustainment are produced concurrently. The potential well also serves as an ion acceleration mechanism, providing essential fuel heating for fusion reactions.

Taken together, these three essential mechanisms enable the Polywell approach to overcome key limitations in standalone high-beta cusp and IEC devices. Unlike conventional IEC systems that rely on physical grids or electrodes, the Polywell magnetically confines electrons, eliminating loss to physical grids, while providing ion acceleration/heating to fusion energies due to the space charge generated by the electron beams. The negative potential well reduces the ion velocity at the plasma edge, which in turn leads to a reduction in ion losses. The reduction of ion loss yields a synergistic improvement in electron confinement within the high beta cusp, due to ambipolar plasma loss constraints. In section 3, we will examine these three essential operating principles of Polywell in detail.

Beyond these three essential mechanisms, Bussard and his colleagues explored additional supplementary concepts to further enhance confinement efficiency and fusion yield in the Polywell, aiming for net energy gain with advanced fusion fuels, such as protons and boron-11 [5,6]. They are:

M4: Spherically convergent ion focus.

Spherically converging ion focus may enhance the fusion yield by significantly increasing the ion density in the central core of the Polywell device, where its kinetic energy is highest due to electrostatic acceleration by the potential well.

M5: Nonthermal plasma operation.

The Polywell may sustain a high degree of nonthermal plasma operation in the core region with an average ion energy significantly higher than the average electron energy. In the case of a proton-boron aneutronic fusion reactor, an example of desirable non-thermal operation parameters is: 300 keV of average ion energy, 20 keV of average electron energy, 320 keV of electron beam energy, and 300 kV of potential well. This level of non-thermal plasma condition would open up the possibility of utilizing proton-boron aneutronic fusion reactions for net energy production, by keeping the energy loss via Bremsstrahlung radiation low, even with high-Z ($Z = 5$) boron ions, while ensuring sufficiently high fusion reactivity. While it may be possible to produce such a deep potential well, a crucial question remains about the power efficiency of sustaining the deep potential well and maintaining non-thermal operation. Bussard and his colleagues proposed that it may be feasible to achieve energetically favorable non-thermal operation by utilizing a spherically converging ion focus and highly efficient electron confinement from a quasi-spherical cusp magnetic field, while avoiding loss of ion focusing and rapid thermalization between ions and electrons [5,6].

M6: Magnetically insulated grids for potential well formation.

As an alternative to electron beam injectors, Bussard proposed forming the potential well by charging the magnet coils, producing magnetically insulated grids. In this concept, high positive voltages are applied to the metal casing of the cusp coils while electron emitters at ground potential provide the excess negative charge for potential well formation. In addition, the positively biased grids may provide an attractive force to facilitate recirculation of escaping electrons back into the magnetic cusp, thus enhancing the electron confinement needed for potential well sustainment.

These supplementary mechanisms were pursued primarily to enable aneutronic fusion and were the major focus of the Polywell research program at Energy Matter Conversion Corporation (EMC2) between 1992 and 2012. However, after years of investigation, EMC2 concluded that the practical implementation of these mechanisms would pose significant technical challenges. Furthermore, they are not essential for achieving net energy gain for the case of conventional deuterium-tritium (D-T) fuels. Consequently, EMC2 suspended R&D on the supplemental mechanisms (M4-M6). Instead, EMC2 refocused its developmental efforts exclusively on the essential mechanisms of Polywell (M1-M3), which offer a tractable path to a compact, net power-producing D-T fusion reactor. Unfortunately, much criticism of the Polywell approach focused on these now-suspended supplemental mechanisms (M4-M6), clouding an objective evaluation of its promise.

In Section 2, we will discuss the rationale that led to the suspension of EMC2's Polywell R&D on the supplementary mechanisms (M4-M6). In Section 3, we will analyze the experimental, theoretical, and simulation results to evaluate the scientific merits of the essential mechanisms (M1-M3). In Section 4, we will discuss the updated Polywell fusion power scaling and provide a preliminary analysis of a Polywell fusion reactor capable of achieving Q > 10, where Q is the ratio between the fusion power output and the input power required to produce and sustain the high-temperature, high-density plasma inside a Polywell device. A discussion and summary will be given in Section 5.

## 2. Suspension of Supplementary Polywell Mechanisms

In this section, we present the experimental observations, theoretical considerations, and strategic rationale that led to the suspension of EMC2's Polywell R&D activities related to supplementary mechanisms M4-M6, as outlined in Section 1. While the arguments may appear self-evident to experienced plasma physicists, we include them here for completeness and to provide historical context. Readers primarily interested in the essential Polywell mechanisms may wish to proceed directly to Section 3.

Figure 1 (a) shows the schematic of the WB-8 (Wiffle Ball-8) Polywell test device that was constructed, operated, and studied at EMC2 under the U.S. Department of Defense Contract No. N68936-09-0125 from 2009 to 2015. The WB-8 was a 6-coil cusp device in a hexahedral (cubic) configuration. Each coil can generate a magnetic field strength of up to 7 kG at its center, with a coil diameter of 40 cm in a cubic vacuum chamber measuring 1.5 meters on each side. WB-8 coil casings were electrically isolated from the vacuum chamber at ground potential up to 50 kV by using insulating rods that also served as mechanical supports. These supports were enclosed in hollow boron nitride cylinders that served as plasma-facing components.

Note that WB-8 was the first Polywell device to implement full magnetic insulation of the high-voltage grids (i.e., the coil casings) such that the entire conductor surfaces facing the plasma were surrounded by magnetic fields parallel to the surface. This configuration effectively prevented plasma loss to the grid. In contrast, the preceding device, WB-7, also a 6-coil cusp system with a 30 cm coil diameter, employed only partial magnetic insulation in an otherwise similar geometric configuration. In WB-7, twelve stainless steel support tubes (0.6 cm in diameter and 1.2 cm in length) were added to the six coil casings. These tubes were located at the centers of the cube edges, with each coil having four support tubes connecting to its adjacent coils. These support tubes mechanically secured the coil assembly against the repulsive forces among the cusp coils. These tubes were not electrically insulated from the coil casings and thus became a part of the grid system in WB-7. Since the support tubes intercepted the magnetic field at the surface, they provided direct pathways for plasma loss to the grids along magnetic field lines, thereby limiting the efficiency of magnetic insulation in WB-7.

The design of WB-8 employed magnetically insulated grids (MIGs) as the primary plasma heating system, consisting of six cusp coil casings where high-voltage power would be applied. In addition, WB-8 could operate with up to 8 electron emitters located at the cube's corners. For clarity, Fig. 1 (a) shows only a single electron emitter at the top-front cusp corner. By applying a high-voltage (HV) positive pulse to the MIGs, the experimental design was that electrons would be accelerated from the emitters (held at or near ground potential) toward the MIGs, providing power for plasma heating and the formation of an electrostatic potential well. The MIGs were powered by a Marx generator capable of delivering up to 2,000 A at voltages of up to 50 kV with a pulse duration up to 10 ms.

The WB-8 device also utilized a coaxial gun-type arc plasma source as a plasma start-up system, with input powers ranging from 20 to 150 kW. The arc source was mounted at the bottom of the chamber. It could be moved vertically as indicated by the purple arrow in Fig. 1(a) to control the plasma injection efficiency and subsequent plasma density in the cusp. It injected cold (3-5 eV) and dense hydrogen plasma ranging from 0.1 to 2 $\times 10^{19}/m^3$ into the central cusp region.

WB-8 was equipped with a suite of plasma diagnostics, including a 94 GHz heterodyne microwave interferometer for line-integrated plasma density measurements across the line of sight passing through the centers of the front and back coils, emissive probes to measure the plasma potential at various spatial location including the center of the cusp coil, x-ray photodiodes and image plates to measure energetic electron behavior associated with potential well formation. Additional diagnostics included magnetic probes measuring plasma diamagnetism and neutron/proton detectors to monitor fusion reactivity.

Fig. 1(b) to 1(d) show plasma emission images from a nominal WB-8 operation with the following experimental conditions: magnetic cusp strength of 2 kG on the center of each coil, plasma initiation using arc source operating at 60 kW, a positive bias of 13.2 kV on the MIGs for a pulse duration of 8.6 ms. These images provide the following two key observations. The brightness distribution in Fig. 1(b) and Fig. 1(c) closely matches the vacuum magnetic field structure as expected in a low beta plasma. Figure 1(d) shows very thin, bright regions located in close proximity to the positively biased coil casings.

Fig. 2 shows the temporal evolution of key parameters during the plasma shot corresponding to the plasma emission images shown in Fig. 1, including the MIG voltage, MIG current, line-averaged electron density by the microwave interferometer, and arc source current for the plasma injection. In the initial phase (0 - 1.5 ms), there is essentially no plasma generation due to the low gas pressure in the chamber (below 0.1 mtorr, chosen to prevent surface flashovers on the MIG structures). At t=1.5 ms, the arc injector is activated with an input power of ~ 60 kW, delivering plasma to the cusp. This triggers a rapid rise in electron density, reaching ~$2.5 \times 10^{18}/m^3$ in the central region of the cusp. During this period (1.5 – 4.5 ms), the MIG current remains low, at ~6 A, corresponding to ~80 kW of power input at ~13 kV. In comparison, the previous Polywell test device, WB-7, operated with a much higher grid current, ~40 A, at the same grid voltage, despite the plasma density being 10-20 times lower in WB-7 compared to WB-8. As such, the use of MIG for WB-8 resulted in more than an order of magnitude reduction in plasma loss to the grid structure, compared to WB-7.

When the arc source is turned off at t=4.5 ms, the electron density in the cusp decays rapidly even though MIG current begins to increase only gradually, accompanied by a modest voltage droop. During this afterglow period from t = 6 ms to 8.5 ms, the plasma density measured by the microwave interferometer drops to negligible levels, indicating the absence of significant plasma in the central region of the cusp, despite continued MIG input power in the range of 130-200 kW.

Fig. 3 shows the temporal evolution of the electron density in the central cusp region before and after termination of the arc source, measured for five different MIG bias voltages. Prior to arc source termination, the electron density is nearly identical for all grid voltages. After the arc source is turned off at 4.13 ms, the electron density decays exponentially with a characteristic time scale of ~180 μs, independent of the MIG voltage.

Fig. 4 shows the temporal evolution of the plasma potential at the center of the top cusp coil, measured by the emissive probe. At t=0, a HV bias of 7.2 kV is applied to the MIGs, with the cusp magnetic field set to 2 kG at the coil center. The emissive probe measures a plasma potential of ~2.2 kV at the center of the top cusp coil, indicating that the formation of an electrostatic potential structure spanning from 7.2 kV at the MIGs, to 2.2 kV at the cusp center, and 0.0 V at the chamber wall. However, once the arc plasma source is turned on at t=1.5 ms and begins injecting plasma into the cusp region, the plasma potential drops rapidly to near ground potential. This near-zero plasma potential persists even after the arc source is turned off at t=4.6 ms until the end of the shot at 6.7 ms, when the MIG bias voltage is terminated. Note that the microwave interferometer data indicate a low but finite residual plasma density, ranging from 0.1 – $1 \times 10^{12}/cm^3$ during this period between 4.6 ms and 6.7 ms.

The following conclusions are drawn from the experimental observations presented in Figures 1 through 4:

i) Insulating grids using magnetic fields are effective in limiting plasma loss to the MIGs in the Polywell cusp configuration. The effectiveness of MIGs results in significantly reduced grid currents in WB-8 compared to WB-7 by more than an order of magnitude when normalized to the plasma density inside the cusp. It is also why the plasma density in the magnetic cusp did not decrease with increasing bias voltage and nearly constant plasma loss rate during afterglow phase was observed independent of grid bias voltages, as shown in Fig. 3. By suppressing the plasma loss to the MIGs, the grid loss became no longer a main factor in the particle balance that determines the plasma density in the cusp.

ii) However, the effective magnetic insulation of grids led to ineffective grid operation for heating and sustaining plasma in the cusp. Without being able to deliver the power to the plasma in the central region of the cusp, biased MIGs could not sustain even modestly dense plasma in the range of 0.1–$1 \times 10^{19}/m^3$ following arc source termination, as shown in Figs. 2 - 4.

iii) The effective magnetic insulation also resulted in highly localized power deposition near the MIG surfaces, as shown in Fig. 1(d). This observation suggests that the electric field generated by the MIG bias was spatially confined to a narrow boundary region adjacent to the grid surfaces, essentially the same as magnetically modified Debye shielding, a well-known phenomenon in plasma physics [7]. Outside this boundary layer, MIGs do not generate a potential well as shown in Fig. 4 with the collapse of the plasma potential structure even at a modest plasma density on the order of $1 \times 10^{18}/m^3$.

Based on the above interpretations, EMC2 suspended the MIG approach (i.e., M6) in 2012, as it became apparent that potential well formation by MIGs at a fusion-relevant plasma density is highly non-trivial and challenging to achieve, even if it may become feasible in the future.

In addition, improved understanding of MIG operation raised a question about utilizing MIGs to control the shape of the potential well to achieve spherical convergence. For example, by using spherically shaped coil casings for MIGs, it was envisioned that an appropriate MIG arrangement may facilitate spherically converging ion focus in the Polywell with the use of a higher-order Polyhedral (e.g., dodecahedron or icosahedron). However, this idea became irrelevant with the suspension of the MIG approach. The observation of a highly non-spherical magnetic field structure in a 6-coil Polywell system, as shown in Fig.1, is another factor that led to the suspension of exploring the spherically convergent ion focus mechanism (M4) in a Polywell. As in the case of MIG, it became apparent that producing and sustaining a spherically convergent ion focus in the Polywell device is highly non-trivial and challenging to achieve, even if it may become feasible in the future.

Lastly, without a clear viable path enabling ion focus, EMC2 decided to suspend its efforts on exploring nonthermal operation (M5) until a later time, as it is more practical to produce net fusion energy in a Polywell device operating in a thermal plasma using D-T fuels. Separately, the authors would like to make the following statements on the subject of nonthermal Polywell operation. The past criticisms by Nevins and Rider against this particular aspect of Polywell fusion have scientific merits based on their arguments on rapid plasma thermalization and velocity space relaxation [8,9]. However, the use of simplistic electric field profiles assumed in the work of Nevins and Rider to model the deficiencies of the Polywell approach to achieve nonthermal operation does not lead to a conclusion that no future variant of Polywell nonthermal operation can achieve net energy generation using advanced fusion fuels such as D-$^3$He or p-$^{11}$B. Consequently, the scientific question of nonthermal operation in Polywell (M5) remains open, but it is no longer the primary focus of the current Polywell D-T fusion effort.

### 3. Analysis of Essential Polywell Mechanisms

Following the challenges of magnetically insulated grids (MIGs) in the WB-8 device, it became evident that energetic electron beam injection is likely the most promising pathway for generating a potential well in the Polywell concept. On the other hand, WB-8 did not have sufficient power to produce a high beta condition, as its design relied on MIGs to couple 100 MW of pulse power to the cusp plasma. In comparison, it had a maximum input power of 150 kW through an arc plasma injector and an additional 30 - 40 kW through electron beam injectors operating at 7 kV. When the electron beams were injected into a low beta cusp plasma in WB-8 with a nominal beta value less than 0.1%, x-ray emission data indicated that the confinement time of 7 keV beam electrons was on the order of 1 μs. With such a poor confinement, the electron beam injection did little to the background cusp plasma, with no signs of plasma heating or potential well formation. This result is consistent with the expected poor confinement of the magnetic cusp system operating at a low beta plasma condition. It also highlights the critical necessity of achieving a high beta plasma condition to facilitate meaningful beam electron confinement, thereby enabling the formation of a potential well.

Before describing the high-beta plasma generation experiment in a subsequent Polywell device, it is worth noting two previous experimental works on high-beta plasma formation in two-coil spindle cusp systems. Between 1974 and 1980, two groups, at the University of Tokyo and the U.S. Naval Research Laboratory, successfully demonstrated the formation of high-beta cusp plasma using high-power laser ablation of deuterium ice pellets. In the work of Kitsunezaki et al., a ruby laser delivering 500 – 800 MW of pulsed power was used to produce a high beta cusp plasma in a spindle cusp with a 3 cm coil diameter and a 3.3 kG magnetic field at the ring cusp location [10]. Similarly, Pechacek et al. utilized a two-stage laser system (a 100 J Nd-glass laser followed by a 1,000J $CO_2$ laser) with input power exceeding 1 GW to generate a high beta plasma in a spindle cusp of 70 cm coil diameter and a 1.7 kG of magnetic field at the ring cusp location [11]. Both high-beta cusp studies highlight the crucial importance of very high power plasma start-up methods in achieving high beta conditions.

In contrast, the maximum power for the plasma start-up in WB-8 was limited to 150 kW from the arc plasma injector. With poor confinement during the initial low beta state, the maximum achievable plasma density was low, as was the plasma pressure. Under a low beta condition, not only was the energetic beam confinement poor, but also the plasma diamagnetism was too weak to expel the vacuum magnetic fields, as shown in Fig. 1 (b) and (c).

To address this challenge, a new Polywell test device, called WB-X, was constructed employing a high-power plasma start-up method that utilizes merging dense plasmoids from two coaxial plasma guns. This new device successfully achieved a high beta condition in a 6-coil Polywell configuration for the first time, and demonstrated enhanced confinement of 7 keV electron beams in a high beta magnetic cusp [12].

Here, we briefly summarize the WB-X device setup and key findings, as the details of the experimental setup and results were previously published and presented in the public forum [12,13].

WB-X was also a 6-coil cusp system, but in a significantly smaller size than WB-8, with a coil diameter of 13.8 cm, compared to 40 cm for WB-8. Similar to WB-8, WB-X utilized independently mounted cusp coils to ensure that coil casings did not intercept the magnetic field lines. On the other hand, coil casings were electrically connected to the vacuum chamber at ground, thus significantly simplifying the device construction. Two coaxial plasma guns utilizing $\vec{j} \times \vec{B}$ plasma acceleration, known as the Marshall gun, provided the plasma start-up. They operated with 60–150 kA of gun current at up to 10 kV, utilizing ignitron switches. Instead of high-pressure gas injection, the coaxial guns utilized solid polypropylene films with a 4 μm thickness to generate high-density plasmas in a short burst between 5-10 μs, with up to 500 MW pulse power. The coaxial guns were mounted on the opposite side of the cusp system to enable two plasmoids to merge into a high-beta plasma inside WB-X during start-up.

Here are the key plasma diagnostics systems employed in WB-X. Temporal evolution of plasma density was measured using a heterodyne laser interferometer at 532 nm, with validation provided by Stark broadening of the H-alpha line using a high-resolution spectrometer. Plasma diamagnetism was measured using flux loops embedded into the coil casings. Hard x-ray emission intensity from the central region of the WB-X from the interaction between injected 7 keV beam electrons and high beta cusp plasma was measured using two spatially collimated x-ray detectors with 25μ thick Kapton filters to block x-rays below 2 keV. Time-resolved plasma emission spectra were measured using an intensified CCD (Princeton Instruments PI-MAX3) to ensure the absence of high-Z impurity that could contaminate the x-ray data during the time window of interest.

Here are the key findings from WB-X experiments [12]:

i) A sharp increase, more than an order of magnitude, in hard x-ray emission above 2 keV was observed after plasma injection at 700 MW input power established a high beta plasma condition, confirming improved confinement of 7 keV electron beams in the high beta 6-coil Polywell cusp, as shown in Fig. 5. Time-resolved experimental results showed that the enhancement confinement phase, green shaded region between 14 μs and 19 μs, began after the peaking of magnetic field exclusion. During this phase, the plasma density remained nearly constant at a high level, 7-8x10$^{21}$/m$^3$, as measured by the laser interferometry. Consequently, the dramatic rise in the hard x-ray emission intensity was attributed to the increased electron beam density from the enhanced confinement inside the cusp in a high beta plasma condition.

ii) The hard x-ray emission intensity was measured at varying plasma injection powers, ranging from 220 MW to 700 MW. Significant electron confinement improvement was only observed when the plasma injection power reached 450 MW or higher, accompanied by corresponding magnetic flux exclusion, reaffirming the necessity of high-power plasma start-up in Polywell high-beta operation.

iii) Experiments conducted at varying cusp magnetic field strengths showed that both too low and too high magnetic fields reduce confinement, indicating the existence of an optimal β value, β ~1.

iv) The observed confinement behavior was consistent with the theoretical projections by Grad and his colleagues at NYU about the enhanced plasma confinement of the high beta cusp, compared to the cusp operating at low beta conditions.

v) While the WB-X experimental results were highly promising, they also highlighted the need to develop a new start-up system for future Polywell experiments with reduced impurity content and improved pulse power coupling to access steady-state, high-beta Polywell operations.

These WB-X experimental results therefore validated the first of three essential mechanisms for Polywell (M1), magnetic cusp confinement of electrons in a high-beta plasma, a critical prerequisite for generating and sustaining a potential well by electron beam injection.

Encouraged by the experimental breakthrough in demonstrating enhanced confinement of high-energy electrons in a high-beta Polywell configuration, EMC2 intensified its efforts to improve the scientific understanding of high-beta plasma dynamics and establish the scientific foundation on the remaining essential mechanisms of the Polywell approach (b) and (c), through advanced computational modeling. This shift in research focus from experiments to computational modeling was also driven, in part, by the funding shortfall following the end of the U.S. DOD contract in 2015.

Here, the authors identify the following key physics processes of high-beta cusp plasma dynamics, which the code needs to accurately capture in its simulations to support the development of the Polywell fusion concept:

P1: The code needs to resolve the zeroth-order exclusion of the externally applied magnetic field from the high-beta plasma volume. This is a defining characteristic of high-beta cusp plasma, where strong diamagnetic currents exclude the vacuum magnetic field from the plasma core. Figure 6 illustrates this phenomenon in a two-coil spindle magnetic cusp configuration. In the low beta case, as shown in Fig. 6 (a), the plasma pressure is insufficient to generate strong diamagnetic currents to exclude the magnetic field from its volume. The cusp plasma co-exists with the magnetic field, and its density varies gradually across the volume. In contrast, in the high beta case, as shown in Fig. 6(b), the plasma pressure is sufficiently high to generate strong diamagnetic currents, which completely exclude the magnetic field from the dense core region, creating a field-free volume surrounded by a narrow boundary layer. As such, accurately capturing this nonlinear reconfiguration of the magnetic field is essential, with self-consistent plasma motion leading to required diamagnetic currents. Satisfying this first requirement was deemed critical in selecting the code.

P2: The code needs to simulate plasma dynamics across two disparate regions in the high-beta plasma state: a field-free core region where a dense plasma is confined with a zero or near-zero magnetic field, and an outer region of low plasma density with strong magnetic fields. These two regions are separated by a narrow boundary layer, where the diamagnetic currents reside and are responsible for excluding the magnetic field. Accurately simulating the plasma dynamics in a wide range of parameter space is essential.

P3: The code needs to simulate the plasma dynamics in the narrow boundary layer, as it plays a crucial role in determining the plasma confinement and loss rate for both electrons and ions. In the field-free core region, the plasma particles move freely. As electrons and ions move outward, they encounter a rapidly increasing magnetic field near the cusp boundary. Many of the particles are reflected back to the core region due to the steep magnetic field gradient in the boundary layer. However, a small but finite fraction attaches its motion to the magnetic field when the gyroradius becomes comparable to the local gradient scale length of the magnetic field in its trajectory. Grad's high beta cusp conjecture is based on suppressing the particle's attachment to the magnetic field line by making the boundary layer very narrow, as shown in Fig. 6 (b), resulting in specular reflection of the particle. In the case of low-beta plasma, however, many of the outgoing particles attach their motion to the magnetic field lines as the magnetic field strength varies only gradually, as shown in Fig. 6 (a). As such, it is essential for the code to simulate particle dynamics accurately and self-consistently in the narrow boundary layer to resolve the process by which the particle attaches its motion to the magnetic field or, conversely, lacks thereof. Note that not all of the particles that attach their motion to the magnetic field lines will escape out of the cusp because of magnetic mirroring in both high-beta and low-beta cases.

P4: The code needs to simulate electric fields in the boundary layer, which influence confinement. Past studies indicated that the plasma confinement of high beta cusp systems would be insufficient for net energy generation if the boundary layer electric field is weak, with its value lower than $E < kT_i/e\rho_i$, where k is the Boltzmann coefficient, $T_i$ the ion temperature, e the elementary charge, and $\rho_i$ the ion gyroradius [4]. In comparison, the high beta cusp system would have sufficiently high confinement for potential net energy generation if the boundary layer electric field is strong, with its value exceeding $E > kT_i/e\rho_i$. As such, the code needs to simulate the strength and profile of the boundary layer electric field, taking into account the self-consistent motion of the plasma across the boundary.

P5: The code needs to simulate ambipolar plasma loss in a self-consistent manner to resolve plasma loss processes in the high-beta cusp plasma. This condition is closely related to condition P4 about the boundary layer electric field. It is also analogous to sheath formation in the presence of electrodes, where self-consistent electric fields arise to equalize the loss rates of ions and electrons at the boundary.

P6: The code needs to simulate complex magnetic field structures of the cusp systems, particularly the 6-coil Polywell configuration, which produces a fully three-dimensional magnetic field structure, as shown in Figs. 1(b) – (d).

P7: Finally, the code needs to simulate the plasma stability of the current layer at the boundary. One of the key merits of the magnetic cusp system is its intrinsic macroscopic plasma stability, resulting from the favorable (concave) curvature of the magnetic field lines toward the confined plasma. On the other hand, the narrow current layer at high beta introduces a steep gradient that can be susceptible to instabilities. Understanding the balance between destabilizing forces from the steep gradient and stabilizing forces from the favorable magnetic field line curvature has been a critical question in past magnetic cusp research [3,4]. As such, it is essential for the code to simulate the stability of the current layer under high-beta conditions.

These physics problems (P1-P7) are extensive and non-trivial. Fortunately, there is a widely studied phenomenon that shares many of the critical physics issues. It is Earth's magnetosphere. The interaction between Earth's magnetic field and incoming solar wind plasmas forms boundary layer currents known as Chapman-Ferraro currents, which play a critical role in preventing unmagnetized solar wind plasma from penetrating the Earth's atmosphere. As a dipole, Earth's magnetic field shares the same topology as the cusp plasma, with the concave field line curvature toward the incoming solar wind plasma. With the interplanetary magnetic field strength between the Sun and Earth being weak, combined with the energetic nature of solar wind plasma, this region mimics the unmagnetized plasma condition of a field-free high-beta cusp region. The magnetopause boundary is where the plasma pressure equals the magnetic field pressure, or the beta = 1 condition. In the inner magnetosphere, the plasma pressure is much weaker than the magnetic field pressure, as in the case of the outer region of high beta cusp plasma. In fact, the review article on cusp confinement research by Spalding highlights the relevance of plasma dynamics in the magnetosphere on the high-beta cusp plasma [3]. As such, attempts were made to utilize well-benchmarked plasma codes for magnetosphere research to simulate the physics of high-beta cusp plasmas and support the development of the Polywell fusion concept.

The first code tried by the authors was a semi-implicit particle-in-cell (PIC) code, called iPIC3D. iPIC3D is an implicit 3D kinetic PIC code optimized for large-scale, multi-scale plasma simulations, developed for modeling of phenomena like reconnection and turbulence in space plasmas, including simulation of the magnetospheres in the solar system [14,15]. iPIC3D is one of the most widely used PIC codes for kinetic simulations of magnetospheres. The upgraded version of iPIC3D, incorporating an energy conservation algorithm, has been successfully employed to conduct the global simulation of Mercury's magnetosphere, treating both electrons and ions as particles in a fully 3D simulation, offering new insights into non-Maxwellian electron distribution from reconnection and kinetic instabilities in Mercury's magnetosphere [16].

Preliminary simulation results of high-beta plasma start-up in a 6-coil Polywell device using iPIC3D were included in a previous study [12]. However, there were significant challenges in achieving numerical stability in the iPIC3D simulation related to the exclusion of the magnetic field by plasma magnetism in a high plasma pressure regime. This issue was challenging to overcome, prompting the use of ECsim (Energy Conserving semi-implicit model), an upgraded version of iPIC3D developed by Lapenta in 2017 [17-19]. The primary difference between the two codes lies in the implementation of the mass matrix algorithm, which enables ECsim to conserve the system's energy precisely, down to machine precision. Unlike iPIC3D, ECsim handled magnetic field exclusion by plasma diamagnetism without any numerical instability in high-beta conditions. The following are the results from the ECsim simulations.

Fig. 7 shows the ECsim simulation results of high beta plasma formation in a 6-coil Polyell cusp system, modeled after previously published experimental results [12]. The simulation utilized 3D Cartesian grids, comprising $100^3$ or $10^{\wedge}6$ cells. The coil diameter was set at 20.0 cm, with a spacing of 32.0 cm between the two opposite coils. The average plasma temperature was ~60 eV, equal for electrons and ions. One caveat of this result was its use of heavy electrons, where the mass of the numerical electron was chosen to be equal to the mass of the numerical ion. This choice allowed for a larger computational time step to resolve gyromotion at the cusp boundary, thereby reducing the required CPU time to obtain quasi-steady-state simulation results.

The results, as shown in Fig. 7, provide good agreement with the theoretical model of high beta cusp plasma by Grad and his colleagues in the key areas: flux exclusion by plasma diamagnetism in the high beta state, boundary layer current generation separating the magnetic field-free dense core region and low density outer region, sharp plasma density gradient at the boundary consistent with favorable plasma confinement in a high beta state [2]. Also, the simulation results highlight the difference in magnetic field line topology. In a low-beta plasma state, magnetic field lines from the outer region are connected to those in the central region, as shown in Fig. 7 (a), providing pathways for plasma loss along these lines. In contrast, field lines from the outer region do not penetrate the central region in a high-beta state, as shown in Fig. 7 (b), thereby suppressing plasma loss along the field line. In Fig. 7 (c), the results also show well-defined boundary layers with diamagnetic currents on the order of 2 $MA/m^2$, where the magnetic field strength rises rapidly from near zero to ~ 0.46T. The result is consistent with the beta ~1 condition at the boundary for the plasma temperature of ~60 eV and the averaged plasma density of ~$4.4 \times 10^{21}/m^3$ in the cusp, as shown in Fig. 7 (d).

Additionally, 3D ECsim simulation results of a 6-coil Polywell device offer insight into the plasma loss in a 6-coil Polywell cusp system at a high-beta state. Fig. 8 shows the plasma loss rate profile using a 3D volume plot. There are six loss regions around the center of each coil, corresponding to the point cusp loss regions in the case of a 2-coil spindle cusp. The observed loss rate in one of the six regions around the center of the coils is ~2.4 MA/m$^2$ over an area of 4.5 cm$^2$, or $\pi(1.2$ cm$)^2$. This value corresponds to ~$n_p V_i^{th} \pi (1.5\rho_i)^2$, where $n_p$ is the average plasma density of $4.4 \times 10^{21}$/m$^3$ in the central region of the cups, $V_i^{th}$ is the thermal velocity of the hydrogen ion, $7.6 \times 10^4$ m/s at 60 eV, and $\rho_i$ is the ion gyroradius of 0.17 cm for 60 eV hydrogen ion at 0.46T of magnetic field. Note that the electron and ion gyroradii are equal in this simulation, with the use of a mass ratio of 1. In addition, there are eight additional loss regions around the corners of the cube whose loss rates are comparable to those of the six point cusp region. On the other hand, there is little plasma loss around the edges of the cube where the directions of the magnetic field are not aligned toward the gradient of the plasma density. The occurrence of eight loss regions at the corners of the cube and the suppressed loss around the edges of the cube are consistent with the published image of the plasma emission profile during high-beta Polywell experiments [12] and anecdotal observations of chamber wall discoloration due to plasma bombardment in both high-beta and low-beta 6-coil Polywell experiments.

This result differs from past studies focused on an axisymmetric 2-coil spindle cusp, where the line cusp loss between two coils was assumed to be dominant and considered as a significant obstacle for net energy generation using cusp devices [2-4,10,11]. Considering that one of the motivations for Bussard to explore 3D polyhedral cusp configurations was to suppress the undesirable line cusp loss in the case of the axisymmetric 2-coil spindle cusp, the result from Fig. 8 is encouraging for two reasons. It shows that a 6-coil Polywell cusp may not suffer from line cusp losses. In exchange, it exhibits a cusp-like loss around the corners, with a loss rate comparable to the point cusps through the center of the coils. While preliminary, it also suggests that the loss rate scales with a small multiple of gyroradius (e.g., 1.5, based on the result from Fig. 8), which could be improved in future devices through optimization. Additional experimental studies and simulations will be necessary to confirm the findings. In the meantime, this encouraging result will be incorporated into the updated Polywell confinement scaling model in Section 4 as part of exploring the parameter space for a Q > 10 Polywell fusion reactor.

While the results from the 3D ECsim simulation were insightful in understanding the high beta cusp plasma dynamics and confinement properties, the numerical resource requirements for full 3D PIC simulations were very high, especially with respect to resolving the boundary layer scale length. In general, PIC simulations require a sufficient number of particles per cell to control noise from particle discreteness, with a nominal number ranging from 100 to 10,000, depending on the specific nature of the relevant physics. Therefore, the required numerical resource would increase by a factor of 1,000 if necessary to increase the spatial resolution by a factor of 10, while maintaining a comparable level of particle noise. Considering that a single 3D 6-coil Polywell simulation using ECsim with only 100 grids in each direction, such as the one shown in Figs. 7 and 8, nominally requires ~ 1 million CPU hours, even with a mass ratio of 1 between electrons and ions; it was necessary to develop a surrogate problem to tackle the Polywell physics. The goal was to explore a wide range of Polywell plasma conditions and conduct quantitative investigations on its confinement properties. Additionally, it was necessary to test and verify the code's capability concerning the physics requirements (P1-P7).

An infinitely long, axisymmetric picket fence with the periodic boundary condition along the axial direction was chosen as a surrogate problem for the Polywell physics. The picket fence was the first magnetic cusp configuration proposed by James Tuck at Los Alamos to address the plasma stability concerns raised by Edward Teller in 1954 regarding interchange instability [20,21]. The picket fence has intrinsic plasma stability due to its favorable magnetic field line curvature and minimum B configuration. By utilizing the axisymmetric picket fence problem, the simulation preserved the favorable magnetic field line curvature and minimum B effects, a critical difference from other studies. For example, the particle simulations of the magnetopause current layer by Berchem and Okuda utilized a minimum B profile, but with a planar geometry, thus excluding the field line curvature effect in plasma stabilization [22]. In the cases of boundary layer studies in beta ~ 1 plasma boundary by Kurshakov and Timofeev [23], and Timofeev, Kurshakov, and Berendeev [24], they utilized uniform magnetic fields across the entire domain, thus excluding both minimum B and field line curvature effects. As such, no attempt will be made to compare our picket fence simulation results with those of the studies mentioned above.

Separately, a 2D version of ECsim (ECsim-CYL), a cylindrical coordinate implementation of ECsim, was developed to enable faster and more efficient PIC simulations for axisymmetric systems such as the picket fence. The ECsim-CYL was tested for its capability to handle energy conservation and its ability to simulate self-consistent ambipolar plasma flow. Here we summarize the key findings from testing of ECsim-CYL, as the details of the code algorithm and results were previously published [25].

i) ECsim-CYL retains the exact energy conservation of the original ECsim with an accuracy of ~ $1 \times 10^{-17}$.
ii) It successfully simulated ambipolar diffusion with charge separation and self-consistent ambipolar electric fields for both reduced and physical mass ratios of 25 and 1836 between electrons and ions. Furthermore, it correctly simulated the energy transfer between electrons and ions via ambipolar electric fields, as validated by the evolution of the electron and ion energy distributions.
iii) It successfully simulated ambipolar diffusion, even with a grid spacing that is 10 times larger than the Debye length, throughout the entire computational domain, confirming the numerical robustness of the semi-implicit algorithm.

After ECsim-CYL was successfully validated, it was utilized to simulate infinitely long, axisymmetric picket fence plasmas that confine high-beta plasma with equal ion and electron temperatures in the central region. Here we summarize some of the key findings from the picket fence simulations, as the details of the code algorithm and results were previously published [26].

i) Comprehensive numerical convergence tests were conducted for ECsim-CYL using a high-beta picket fence problem to examine grid resolution, time step size, mass ratio, and number of particles per cell. In all cases, converged solutions were obtained to validate code outputs and gain scientific insights into the high beta cusp plasma dynamics.
ii) As in the case of 3D 6-coil ECsim simulations, ECsim-CYL was able to simulate magnetic field exclusion due to plasma diamagnetism, plasma dynamics across the entire cusp system, covering both the field-free high-beta central region and the low-density outer region, and a realistic magnetic cusp geometry of the picket fence system, satisfying the code requirements of P1, P2, and P6.
iii) Both ambipolar plasma loss and self-consistent boundary layer electric fields were properly simulated, with successful convergence tests on grid resolutions as low as 0.3 times the electron gyroradius, satisfying the code requirements of P3, P4, and P5.
iv) The results showed that the boundary layer thickness is between 1 and 2 times the electron gyroradius for both the diamagnetic current layer and the localized ambipolar electric layer.

The role of the boundary electric field was demonstrated in reducing ion loss in the high beta plasma state with the electric field amplitude on the order of $E \sim 0.5\ kT/e\rho_e$, where T is the plasma temperature, and $\rho_e$ is the electron gyroradius. This electric field amplitude is significantly greater than the previously theorized cusp electric field threshold of $E = kT_i/e\rho_i$, thus consistent with the observed sharp boundary and steep density gradient, indicating a high level of plasma confinement.

These results, if reproduced in future experiments and simulations, are encouraging and could establish a strong scientific foundation for high-beta cusp systems and Polywell as a potential fusion approach. Here we present two additional results from these picket fence simulations to strengthen the case of the Polywell fusion approach.

Fig. 9 shows the electric field profile and magnetic field strength in the high-beta picket fence from the high-resolution simulation, i.e., the case of Run 1 in the paper by Park et al [26]. It employed a grid resolution of 0.3 times the electron gyroradius at the cusp boundary with $1.76 \times 10^8$ numerical particles in the entire 2D domain of 540 x 360 cells and a nominal particle density of ~ 1,000 electrons and ions per cell at the boundary region with a mass ratio of $(m_i/m_e) = 64$. As highlighted by the black dotted box in Fig. 9 (a), the electric fields in the outer region of the picket fence show a clear sign of plasma instability with the electric field changing its direction on a scale of a few electron gyroradii. In comparison, no sign of plasma instability was observed inside the picket fence despite the presence of a narrow boundary layer and steep gradients in plasma density and diamagnetic current. Here, we note potential causes for the occurrence of plasma instability in the outer region and the absence of instability in the inner region. The first possible explanation is that the magnetic field line curvature in the outer region is significantly smaller than in the inner region, thus providing an insufficient stabilizing force against destabilizing forces from the steep gradients. In comparison, the field line curvature is significant in the inner region. The second is that this unstable region is adjacent to the region of finite B-field where the plasma is magnetized and can support a wide range of instabilities to grow. In comparison, the inner region is adjacent to the region of near-zero B-field where the plasma is unmagnetized, thus limiting the type of instabilities. While the results shown in Fig. 9 provide the validation that ECsim-CYL is capable of simulating some plasma instabilities (P7), the authors acknowledge that the topic of plasma instability requires significant future work, both in experiments and theoretical modeling. Nonetheless, it is encouraging that no sign of plasma instability was observed inside the high beta picket fence in ECim-CYL simulations, as expected due to the favorable field line curvature.

The ability to control the mass ratio is a useful feature in ECsim and ECsim-CYL [25,26]. It shortens the simulation time by using reduced mass ratios, such as 64, 256, or even 1. It also offers insights into the dynamics of ambipolar plasma loss by parametrically controlling the mass ratio. Note that in ECsim and ECsim-CYL, typically the electron mass is adjusted rather than the ion mass to vary the mass ratio. For example, a mass ratio of 1 is the case where the electron mass is the same as the proton mass, and 64 is the case where the electron mass is one $64^{th}$ of the proton mass. A particularly interesting case would be a mass ratio of 1, which eliminates critical differences in electron and ion motion related to their thermal velocities and gyroradii. In this case, there would be no ambipolar electric field, as shown in previous work [26], as well as the 3D ECsim results shown in Figs. 7 and 8. With an increase in the mass ratio, the ambipolar electric field arises self-consistently to adjust the plasma loss rate by slowing down the loss rate of the species with the faster intrinsic loss rate in the absence of an ambipolar electric field. Note that ions, and not electrons, would have a higher loss rate in the high beta cusp in the absence of an ambipolar electric field due to their large gyroradius at the cusp boundary. This is because a large gyroradius leads to an increase in the probability of attaching its motion to the magnetic field. Note that the sign of the ambipolar electric field shown in Fig. 9 (a) is to slow the ion motion and its loss at the cusp boundary.

Fig. 10 shows the plasma loss rate of high beta picket fence as a function of the mass ratio between numerical ions and electrons from ECsim-CYL simulations [26]. The loss rate decreases with increasing mass ratio (i.e., decreasing numerical electron mass in the simulation), indicating the role of the electric field and its strength in reducing the ion loss rate in the high-beta picket fence cusp system. The observed reduction factor has an exponent factor of 0.28, which can be described as the hybrid gyroradius scale.

In their work in the 1950s, Grad and his colleagues formulated the plasma loss rate of the 2-coil spindle cusp in a high-beta state, $L_{loss}^{spindle}$, using the following expression.

$$L_{spindle} \propto n_p V_{ion} <L_{cusp}> 2\pi R \quad (1),$$

where $n_p$ is the plasma density in the cusp, $V_{ion}$ is the ion thermal velocity, and $2\pi R$ is the circumference of the line cusp region with a coil radius of R. In addition, Eq. 1 utilizes $<L_{cusp}>$, a factor determining the cusp loss rate, which would have the unit of length and could be calculated if one can properly account for particle velocity distribution and its motion at the boundary along with the presence of self-consistent boundary layer electric field that affects the probability of particle attaching its motion to the magnetic field [2]. In the case of an optimistic projection by Grad, $<L_{cusp}>$ would scale as the electron gyroradius, enabling a net energy-producing cusp fusion reactor. In comparison, other studies of high beta cusp indicated that $<L_{cusp}>$ would scale as either ion gyroradius or hybrid gyroradius [3,4,10,11]. The high beta cusp systems would not be capable of generating net fusion energy if $<L_{cusp}>$ scales with ion gyroradius. The results in Fig. 10 show that $<L_{cusp}>$ in the case of a high beta picket fence would scale close to the hybrid gyroradius with an exponent of 0.28. In comparison, the exponent would have been 0.5 if $<L_{cusp}>$ scales with the electron gyroradius or 0.0 with the ion gyroradius.

As previously discussed in Fig. 8, a 6-coil Polywell cusp loss rate is not dominated by the line cusp loss. Instead, its particle loss rate may be expressed in the following form utilizing the $<L_{cusp}>$ factor introduced by Grad in Eq. 1.

$$L_{6\text{-coil}} \sim 14 n_p V_{ion} \pi <L^*_{cusp}>^2 \qquad (2),$$

where $L_{6\text{-coil}}$ is the particle loss rate of the 6-coil cusp system in the high beta state, 14 is the sum of 6 point cusps and 8 point-line cusps for the corners of the cube, and $<L^*_{cusp}>$ is the factor determining the cusp loss rate for the 6-coil Polywell cusp. The result from Fig. 8 indicates that $<L^*_{cusp}> \sim 1.5\rho_i = 1.5\rho_e = 1.5\rho_{hybrid}$, since the mass ratio of 1 was utilized in the simulation. Now a question arises about the value of $<L^*_{cusp}>$ in a real physical system where the mass ratio of Deuterium ions and electrons is 3672. While preliminary, the authors believe that the likely value would be $<L^*_{cusp}> = 1.5\rho_{hybrid}$. This hypothesis is based on the role of ambipolar electric fields that must arise to suppress the ion loss rate in the high-beta cusp system, as shown in Fig. 10. Note that the $\rho_i^2$ is much larger than $\rho_{hybrid}^2$ by a factor of 60.6 for deuterium ions, a significant factor that needs to be validated in future experiments and/or simulations.

While the loss rate in Eq. 2 is encouraging, it remains economically non-trivial to build a power plant based on the conventional high beta cusp system, even if $<L_{cusp}>$ scales with the hybrid gyroradius. This point will be discussed in Section 4. This is where Bussard's idea of adding a potential well to the high-beta cusp becomes critical in the success of the Polywell fusion approach.

Fig. 11 shows the ECsim-CYL results, comparing the high-beta plasma properties in a picket fence without and with electron beam injection. With electron beam injection, the plasma inside the picket fence develops an additional negative space charge, thereby strengthening the boundary layer electric field. This increase in electric field strength beyond the ambipolar electric field in the absence of electron beam injection results in significantly reduced plasma density by a factor of 10 or so in the outer region, as highlighted by the yellow dotted boxes. This result illustrates the principal dynamics of the essential Polywell mechanisms M2 and M3 that can lead to improved plasma confinement with electron beam injection and potential well formation. While the results shown in Fig. 11 are encouraging, it is noted that simulating electron beam injection was numerically challenging in terms of achieving a quasi-steady state in the simulation, unlike other simulation results presented previously [26] and in this work. This numerical challenge is in large part attributed to the evolution of strong downstream electric field instabilities, as shown in the top right panel of Fig. 11 (b). This region caused a breakdown in ECsim-CYL's energy conservation algorithms, resulting in numerical instabilities that hindered code convergence. The work is currently ongoing to address this issue, along with other numerical challenges related to revisiting a full 3D 6-coil Polywell simulation.

In summary, EMC2 has conducted a systematic investigation into testing and validating the essential Polywell fusion mechanisms through experiments and numerical simulations. The work to date has shown the following:

i) Mechanism 1 (M1) was demonstrated in WB-X experiments.
ii) 3D 6-coil Polywell simulation indicates that it may be possible to avoid the undesirable line cusp loss with the use of 3D polyhedral cusp configurations.
iii) Extensive numerical investigation of high beta cusp plasma dynamics was conducted with the use of an infinitely long axisymmetric picket fence cusp configuration with the periodic boundary condition along the axis. The results provided validation of the code's capability and a strong scientific foundation for the high-beta cusp confinement and Polywell fusion approach.
iv) One of the key findings from the picket fence simulation is the role of ambipolar electric fields in regulating plasma loss, with a characteristic length scale of the hybrid gyroradius. Combined with the 3D 6-coil simulation results on the high beta cusp loss rate, the hybrid gyroradius scaling due to ambipolar electric fields offers an encouraging pathway for Polywell fusion.

v)  While preliminary, the electron beam injection simulation results in Fig. 11 illustrate the working principles for the other two essential Polywell mechanisms, M2 and M3, that can lead to synergistic confinement enhancement with the electron beam injection.

vi) Simulation results and anecdotal experimental observations indicate that the high-beta cusp plasma would be stable against both macroscopic and microscopic instabilities due to its favorable field line curvature and minimum B configuration. Additionally, the plasma remains unmagnetized within the majority of the system volume, thereby limiting the types of instabilities.

These results are encouraging and offer a promising path toward overcoming confinement losses and achieving a net energy gain for a Polywell fusion device, as will be discussed in Section 4.

## 4. Polywell Fusion Scaling for Net Energy Generation

Based on the findings from Section 3, we describe the scaling of Polywell fusion and outline a path to net energy generation. We will begin with a plasma loss model for a 6-coil Polywell cusp incorporating the results from 3D ECsim simulations in the absence of electron beam injection based on Eq. 2.

Polywell power loss without potential well (in Watts):

$$P^L_{\text{no-well}} = 14\, n_p v_{ion} (<E_{ion}> + <E_e>) \pi <L^*_{cusp}>^2 \qquad (3),$$

where $<E_{ion}>$ and $<E_e>$ are the average energy of electrons and ions in the cusp, and $<L^*_{cusp}>^2$ is the cusp loss factor for a 6-coil Polywell cusp with a unit of surface area. We will then utilize the hybrid loss rate scaling from the picket fence, as shown in Fig. 10, by setting $<L^*_{cusp}> = 1.5\sqrt{\rho_e \rho_i}$. Note that a multiplication factor of 1.5 was retained from the results shown in Fig. 8, though there is anticipation that future Polywell development could improve the plasma confinement performance and reduce this factor. We will then assume that $<E_{ion}> = <E_e> = kT_{plasma}$ based on a sufficiently fast thermalization time scale for a high-density cusp plasma equilibrium, which will be verified later.

Eq. (3), combined with the selections of $<L^*_{cusp}>$, $<E_{ion}>$, and $<E_e>$, allows us to evaluate the Polywell confinement time scaling as a function of cusp magnetic field strength. Note that the Polywell system must maintain a high beta state during changes in magnetic field or plasma temperature, which governs the relationship among magnetic field, plasma density, and temperature.

We will begin with the change in magnetic field strength by expressing $B = \alpha B_0$, where $\alpha$ is a parametric variable for the change in cusp magnetic field strength and $B_0$ is the reference magnetic field strength. When there is an increase in B-field, the plasma pressure inside the Polywell cusp must increase, by increasing plasma density, its temperature, or both. This can be achieved in operation with increased plasma fueling and/or plasma heating to maintain the high beta equilibrium state, by employing a range of operation scenarios specific to the experimental setup.

First, we will examine the case (Case 1) when the plasma temperature is increased linearly with the magnetic field, corresponding to $T_1 = (B/B_0)T_0$, where $T_1$ is the increased new equilibrium plasma temperature at B and $T_0$ is the reference temperature at $B = B_0$. Due to the high beta constraint, there must be increased plasma fueling as well to increase $n_1 = (B/B_0)n_0$, where $n_1$ is the increased plasma density and $n_0$ is the reference density at $B = B_0$. Eq. 4 summarizes the parametric expressions among B, T, and n, when the Polywell plasma temperature increases linearly with the magnetic field with increased plasma heating.

$$B = \alpha B_0, T_1 = \alpha T_0, \text{ and } n_1 = \alpha n_0 \qquad (4).$$

In this case, we can obtain the following scaling for Polywell cusp confinement time related to the change in magnetic field in the absence of a potential well, as shown in Eq. (5).

$$\tau_{1,no-well} = \frac{W_{no-well}}{P^L_{no-well}} = \frac{\alpha^2}{\alpha^{1.5}} \tau_0 = \left(\frac{B}{B_0}\right)^{0.5} \tau_0 \qquad (5),$$

where $\tau_{1,no\text{-}well}$ is the Polyell cusp confinement time for Case 1 with increasing B-field, $W_{no\text{-}well}$ is the stored energy inside the 6-coil Polywell cusp, and $\tau_0$ is the reference confinement time at $B=B_0$. As such, the improvement in Polywell confinement time in Case 1 would be limited to $(B/B_0)^{0.5}$ with the increase in magnetic field.

In comparison, it may also be possible to operate a Polywell device in a high beta equilibrium by significantly increasing fueling, more so than in Case 1, while increasing the magnetic field.

In this case (Case 2), we set the density increase to be $n_2 = (B/B_0)^{1.5}n_0$. Eq. 6 summarizes the parametric expressions among B, n, and T for Case 2, and the corresponding scaling in confinement time.

$$B = \alpha B_0, n_1 = \alpha^{1.5}n_0, and\ T_0 = \alpha^{0.5}T_0 \rightarrow \tau_{2,no-well} = \left(\frac{B}{B_0}\right)^{1.25}\tau_0 \qquad (6),$$

where $\tau_{2,no\text{-}well}$ is Polyell cusp confinement time for Case 2, when the plasma density increase is given by $(B/B_0)^{1.5}$. In Case 2, the improvement in Polywell confinement time would be by $(B/B_0)^{1.25}$, with the increase in magnetic field. The results from Eqs. (5) and (6) therefore show that there is a need to explore various operating scenarios regarding plasma fueling and heating to generate experimental data to examine the Polywell scaling and to optimize plasma performance toward a net energy-producing device.

We will now rewrite Eq. 3 in the following form.

$$P_L^{no\text{-}well} = (<E_{ion}> + <E_e>)I_{no\text{-}well} \qquad (7),$$

where $I_{no\text{-}well}$ is the loss rate of ions without a potential well in units of particles/s or Ampere (if combined with the use of eV for $<E_{ion}>$ and $<E_e>$), and can be estimated using Eq. 3. For example, for a 6-coil Polywell device in a 1.6 m length cube, a 4.5T cusp magnetic field strength at the boundary, and operating with a plasma temperature of 20 keV and density of $1.3 \times 10^{21}/m^3$, the loss current without the potential well would be 6.6 kA. This loss rate leads to 265 MW of plasma power loss, since $(<E_{ion}> + <E_e>) = 2kT_{plasma}$, compared to a stored energy of 33 MJ (assuming a plasma volume of 4.1 $m^3$). While still respectable at 0.12 s with a corresponding $nT\tau$ value of $3.1 \times 10^{21}$ keV s/$m^3$, the resulting confinement time would still be short for practical net energy generation. Note that the collisional times for electrons and ions at 20 keV are 77 μs and 6.6 ms, respectively, in a plasma density of $1.3 \times 10^{21}/m^3$, sufficiently short compared to the confinement time to ensure thermalization.

One of the key hypotheses of the Polywell fusion approach is the reduction in the ion loss rate due to a potential created by electron beam injection, which strengthens the boundary layer electric field, as shown in Fig. 11. Currently, we lack a quantitative model for the reduction in the loss rate. Therefore, we will use a parametric expression to represent this reduction in the energy loss rate in this paper.

$$I_{well} = \gamma I_{no\text{-}well} \qquad (8),$$

where $\gamma$ is the loss reduction factor between 0 and 1 as a function of electron beam power and the boundary layer electric field strength that determines the cusp loss factor, $<L^*_{cusp}>$, due to the change in ion velocity and its gyroradius. In the case of efficient loss reduction by a potential well, we can assume the following relationship among the potential well depth, the average ion and electron energies in the central region of the Polywell cusp.

$$V_{well} \sim <E_{ion}> \sim <E_e> \qquad (9),$$

where $V_{well}$ is the potential well depth with the electron beam injection. Note that beam electrons would also thermalize quickly inside the Polywell cusp from frequent collisions at a high plasma density. With $V_{well}$ nearly equal to $<E_{ion}>$, most ions would slow down and return to the central region due to the electric field at the boundary layer. Only a small fraction of ions would overcome the potential and escape the cusp system. On the other hand, escaping electrons would gain energy from the electric field and leave the system with the average energy twice the potential well depth. Therefore, we can utilize Eqs. (7) and (8), to quantify the power loss from the Polywell device with the potential well in a steady state, by utilizing the charge balance constraint for no net charge accumulation.

$$P_L^{well} = 2V_{well}(I_{beam} + I_{well}) \text{ or } 2V_{well}(I_{beam} + \gamma I_{no\text{-}well}) \quad (10),$$

where $P_L^{well}$ is the Polywell power loss with a potential well (in Watts) and $I_{beam}$ is the electron beam injection current for the potential well formation and sustainment.

In comparison, the input power from the electron beam injection would be given as

$$\text{Input Power} = E_{beam}I_{beam} \quad (11),$$

where $E_{beam}$ is the electron beam energy. In a steady state, input power and power loss must be balanced, resulting in the following equation that relates the potential well depth, $V_{well}$, to the electron beam injection energy, $E_{beam}$, and the loss reduction parameter, $\gamma$.

$$V_{well} = 0.5 E_{beam} \frac{I_{beam}}{I_{beam} + \gamma I_{no-well}} \quad (12).$$

Eq. (12) can be utilized to estimate the required electron beam power to produce a potential well depth suitable for fusion power generation as a function of potential well loss reduction efficiency $\gamma$. For example, a potential well of 20 keV can be formed using a 60 keV electron beam injection energy with an injection current of 1.3 kA, assuming this Polywell device achieves a loss reduction efficiency of 0.1. In this case, the input power for this Polywell device would be 78 MW. Note that the input power requirement would increase or decrease to 156 MW or 39 MW for the loss reduction factor of 0.2 or 0.05. In comparison, the estimated fusion power output of this Polywell device with 1.6 m in cube length, operating at a 20 keV plasma temperature and a 4.5T cusp magnetic field at the boundary with a 50:50 D-T fuel mixture, would be approximately 980 MW with a fusion reactivity of $<\sigma\upsilon> \sim 2.2\times10^{-22}$ m$^3$/s, resulting in a Q value of 10.5 for the loss reduction factor of 0.1 and the bremsttrahlung power loss of 15.5 MW.

Here, we summarize key assumptions made to derive the Polywell fusion power scaling, as shown in Eqs. (3) - (12), and illustrate a set of reactor parameters for a compact scale, Q=10 Polywell D-T fusion reactor.

1. The value of the point cusp loss factor $<L^*_{cusp}> = 1.5\sqrt{\rho_e \rho_i}$ in a 6-coil Polywell cusp in Eq. 3 is based on the applicability of the 2D picket fence simulation results to the 3D 6-coil Polywell cusp system with respect to the role of ambipolar electric fields in regulating the plasma loss.
2. Detailed physics of the Polywell mechanisms (b) and (c), which relate to electron beam injection, potential well generation, and synergistic plasma loss reduction, was greatly simplified using a free parameter $\gamma$, the loss reduction factor. A value of 0.1 for $\gamma$ was chosen based on the qualitative interpretation of the PIC simulation results shown in Fig. 11, where a visibly pronounced decrease in ion and electron density in the outer region, an indication of particle loss reduction, is observed with electron beam injection and an increased boundary layer electric field strength.
3. The plasma stability, both against macroscopic and microscopic instabilities, was assumed based on the PIC simulation results shown in Figs. 9 and 11, which show no sign of plasma instability in the confined cusp plasma region.
4. Details of energy exchange among ions, electrons, and injected beam electrons inside the cusp confined region were neglected based on expected rapid thermalization among these species at high plasma densities on the order of $10^{21}$/m$^3$. As a result, a single value of average energy was used for all three species.

## 5. Summary and Discussion

The Polywell fusion concept, originally proposed by Robert W. Bussard in 1985, combines high-beta magnetic cusp confinement of electrons with electrostatic ion confinement through an electron-beam-generated potential well. Despite its promising premise, historical progress has been impeded by inadequate plasma confinement during start-up. In this work, we revisited and examined the essential Polywell mechanisms related to the plasma confinement, utilizing experimental results from the WB-8 and WB-X devices, as well as first-principles PIC simulations with the ECsim code. The study shows a credible path for achieving net fusion energy based on the updated Polywell scaling model.

Authors would like to point out that the assumptions utilized in deriving the Polywell scaling model in Section 4 can be tested and verified in future experiments on a modest-scale device comparable to WB-8, leveraging recent progress in fusion energy technology. For example, the high-power plasma start-up system can utilize recent advances in FRC plasma research to produce and translate dense and hot plasmoids into the cusp system, thereby eliminating past problems in WB-X related to impurity and pulse power coupling. In addition, commercial-grade MW-class electron beam injectors are available that can provide sufficient power to produce and sustain a potential well on the order of 2-4 keV based on Polywell scaling. Recent advances in x-ray tomography systems can be used to measure the generation and sustainment of this level of potential well by measuring not only the x-ray emission intensity but also the anticipated downward energy shift in the x-ray spectrum in the presence of a potential well. Most importantly, future Polywell experiments can now take advantage of the significant scientific progress made by first-principles PIC simulations in their design and operation, with well-defined physics tasks to examine their potential for net fusion energy generation. We note that while the present scaling model has several optimistic projections, a reduction in confinement time of up to a factor of 10 can be compensated by increasing the reactor size and/or magnetic field strength. Additionally, approaches such as employing RF waves to control the particle velocity space distribution at the boundary layer may offer enhanced confinement.

Furthermore, significant progress has been made in ECsim's capability to assist Polywell fusion development in recent years [19]. In particular, a successful implementation of ECsim in curvilinear coordinates offers a potential breakthrough in tackling Polywell's boundary layer problems related to spatial resolution and numerical instability by employing spatially non-uniform grids [27]. Additionally, a first GPU version of ECsim was recently completed, offering ~5x improvement in code speed with ongoing further optimization [28]. Combined with the continued progress in high-performance computing capability and the advent of machine learning and AI, these advances in code capability will be crucial in the timely development of Polywell fusion, complemented by experimental efforts to strengthen its scientific foundation.

We also highlighted significant engineering and economic advantages of the Polywell fusion reactor, which continues to motivate ongoing research. As a high-beta approach, it can deliver substantial fusion power density in a compact-sized device. Its primary heating system, electron beam injection, is a mature technology with off-the-shelf availability of steady-state electron beam injectors in a compact footprint. It also has excellent intrinsic plasma stability for operation reliability to deliver a high facility availability factor for a power plant. A Polywell reactor utilizes compact, non-interlocking coils that can be easily assembled and disassembled in a modular manner, thereby simplifying its deployment and maintenance. Naturally diverging magnetic fields at plasma-facing surfaces facilitate effective thermal management of plasma exhaust. Additionally, tritium breeding blankets can operate in regions of low magnetic field strength, providing opportunities for innovative breeding solutions to address neutron shadowing caused by internal coil structures. Collectively, these positive attributes strongly support further development of Polywell fusion as a practical, scalable, and economically viable approach to fusion energy, reinforced by the newly updated scientific foundation presented herein.

**Acknowledgements**:

We are grateful to the family of the late Prof. Giovanni Lapenta for granting permission to include him as a co-author. The experimental work on WB-8 and WB-X was performed under Contract No. N68936-09-0125, awarded by the U.S. Department of Defense with contributions from former EMC2 team members Michael Skillicorn, Paul E. Sieck, Dustin T. Offermann, Kevin Davis, Andrew Sanchez, Eric Alderson, Mike Wray, Kevin Wray, and Noli Casama. The authors would like to thank Dr. Alan Roberts and Mr. David

T. Mansfield of EMC2, and Mr. John Daly of IonQ, Inc., for their assistance in editing the manuscript. A portion of the funding for this work was provided by the internal corporate research and development program of Energy Matter Conversion Corporation (EMC2). The simulation results were produced with the computational resources provided by NASA Advanced Supercomputing (NAS) and NASA Center for Climate Simulation (NCCS) facilities, the Flemish Supercomputing Center (VSC), and by PRACE Tier-0 allocations.

**Competing Interests**:

The authors have no competing interests to declare that are relevant to the content of this article.

**Author Contributions:**

Jaeyoung Park contributed to the development of Polywell fusion power scaling, incorporating past experimental and computational results, the scaling utilization to characterize a near-term physics validation device and a net energy producing Polywell system, conducted ECsim simulations of Polywell devices, and prepared the manuscript. Nicholas A. Krall contributed to the development of the theoretical framework and analysis of Polywell fusion power scaling and edited the manuscript. Giovanni Lapenta contributed to the customization of ECsim code for Polywell device simulations, conducted ECsim simulations of Polywell devices, and analyzed simulation results. Masayuki Ono contributed to the analysis of Polywell fusion power scaling and development of plasma diagnostics strategy for a near-term physics validation device, and edited the manuscript.

**References:**

1. R. W. Bussard, Method and Apparatus for Controlling Charged Particles, U.S. Patent No. 4,826,646A, Filed on Oct. 29, 1985, Issued May 2, 1989.
2. J. Berkowitz, K. O. Friedrichs, H. Goertzel, H. Grad, J. Killeen, and E. Rubin, Proceedings of the 2nd United Nations International Conference on Peaceful Uses of Atomic Energy Vol. **31**, 171-197 (United Nations, Geneva, 1958).
3. I. Spalding, Cusp Containment, *Advances in Plasma Physics Vol. 4*, edited by A. Simon, W. B. Thomson, p79-123 (Interscience, New York, 1971).
4. M. G. Haines, *Nucl. Fusion* **17**, 811 (1977), DOI: 10.1088/0029-5515/17/4/015.
5. R. W. Bussard, *Fusion Sci. Technol*. **19**, 273 (1991), DOI: 10.13182/FST91-A29364.
6. N. A. Krall, *Fusion Sci. Technol*. **22**, 42 (1992), DOI: 10.13182/FST92-A30052.
7. P. C. Stangeby, *The Plasma Boundary of Magnetic Fusion Devices*, CRC Press (2000), DOI: 10.1201/9781420033328.
8. W. M. Nevins, *Phys. Plasmas* **2**, 3804 (1995), DOI: 10.1063/1.871080.
9. T. H. Rider, *Phys. Plasmas* **4**, 1039 (1997), DOI: 10.1063/1.872556.

10. A. Kitsunezaki, M. Tanimoto, and T. Sekiguchi, *Phys. Fluids* **17**, 1895 (1974), DOI: 10.1063/1.1694636.

11. R. E. Pechacek, J. R. Greig, M. Raleigh, D. W. Koopman, and A. W. DeSilva, *Phys. Rev. Lett*. **45**, 256 (1980), DOI: 10.1103/PhysRevLett.45.256.

12. J. Park et al, *Phys. Rev. X 5*, 021024 (2015), DOI: 10.1103/PhysRevX.5.021024.

13. J. Park, *Polywell Fusion: Electrostatic Fusion in a Magnetic Cusp*, Microsoft Research, Jan. 22, 2015, https://www.microsoft.com/en-us/research/video/polywell-fusion-electrostatic-fusion-in-a-magnetic-cusp/.

14. S. Markidis, G. Lapenta, and R. Uddin, *Mathematics and Computers in Simulation*, **80**, 1509 (2010), DOI: 10.1016/j.matcom.2009.08.038.

15. G. Lapenta, *J. Comput. Phys*. **231**, 795–821 (2012), DOI 10.1016/j.jcp.2011.03.035.

16. G. Lapenta et al., *J. Geophys. Res.: Space Physics* **127**, e2021JA030241 (2022), DOI: 10.1029/2021JA030241.

17. G. Lapenta, Exactly energy conserving semi-implicit particle in cell formulation, *J. Comput. Phys*. **334**, 349 (2017), DOI: 10.1016/j.jcp.2017.01.002.

18. D. Gonzalez-Herrero, E. Boella, and G. Lapenta, *Comput. Phys. Comm*. **229**, 162 (2018), DOI: /10.1016/j.cpc.2018.03.020.

19. G. Lapenta, *Physics* 5, 72 (2023), DOI: 10.3390/physics5010007.

20. A. S. Bishop, *Project Sherwood: The U. S. Program in Controlled Fusion*, Addison-Wesley, Reading, MA, (1958).

21. J. L. Tuck, Washington Report No. 184, *Plasma Physics and Thermonuclear Research Vol. 2 (Progress in Thermonuclear Research)*, edited by C. L. Longmire, J. L. Tuck, and W. B. Thompson (Pergamon Press, London, 1963).

22. J. Berchem, H. Okuda, *J. Geophys. Res* **95**, 8133-8147 (1990), DOI: 10.1029/JA095iA06p08133.

23. V. A. Kurshakov and I. V. Timofeev, *Phys. Plasmas* **30**, 092513 (2023), DOI: 10.1063/5.0153855.

24. I. V. Timofeev, V. A. Kurshakov, and E. A. Berendeev, *Phys. Plasmas* **31**, 082512 (2024), DOI: 10.1063/5.0216073.

25. D. Gonzalez-Herrero, A. Micera, E. Boella, J. Park, and G. Lapenta, *Comput. Phys. Comm*. **236,** 153 (2019), DOI: doi.org/10.1016/j.cpc.2018.10.026.

26. J. Park, G. Lapenta, D. Gonzalez-Herrero, and N. A. Krall, *Front. Astron. Space Sci*. **6**, 74 (2019), DOI: 10.3389/fspas.2019.00074.

27. J. Croonen, L. Pezzini, F. Bacchini, and G. Lapenta, *ApJS* **271**, 63 (2024), DOI: 10.3847/1538-4365/ad31a3.

28. N. Shukla, E. Boella, F. Spiga, M. Redenti, M. K. Chimeh, and M. E. Innocenti, *Optimizing the ECsim Plasma Code for Exascale Architectures: GPU Acceleration, Portability, and Scalability*, poster presentation at The Platform for Advanced Scientific Computing (June 2025, Brugg, Switzerland).

**Figures**

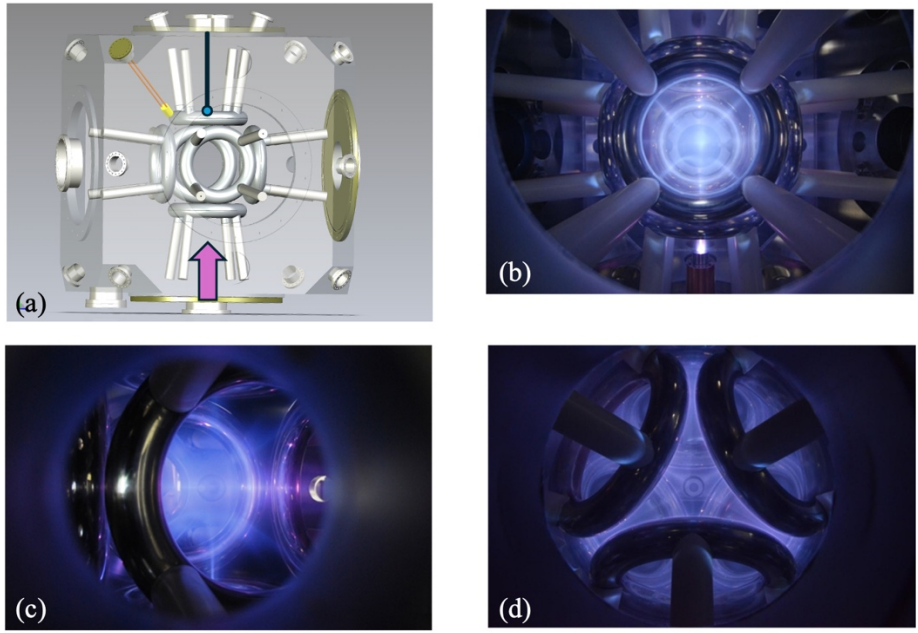

**Fig. 1** (a) A schematic of WB-8 Polywell test device showing six independently mounted cusp coils, an electron emitter (top left), an arc plasma injector (bottom) and an emissive probe for plasma potential measurement (top), and (b) – (d) photographic images of plasma emission from various viewing directions showing a three-dimensional spatial structure of cusp plasma operating at a low beta condition, with the plasma injection from the arc plasma injector at the bottom of the chamber

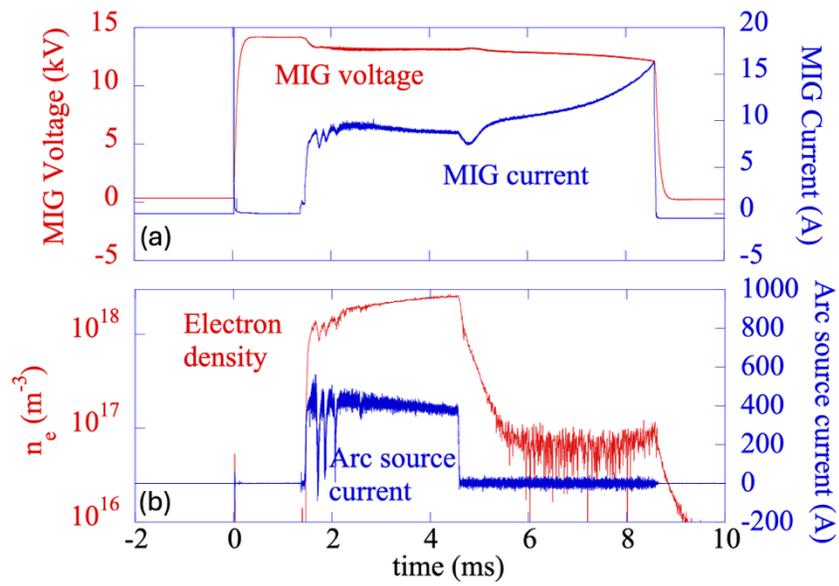

**Fig. 2** Nominal WB-8 plasma operation characteristics for the shot shown in Fig. 1. (a) Temporal variation of magnetically insulated grid (MIG) voltage and MIG current. (b) Temporal variation of line-averaged electron density in the cusp, and arc source current for the plasma injection

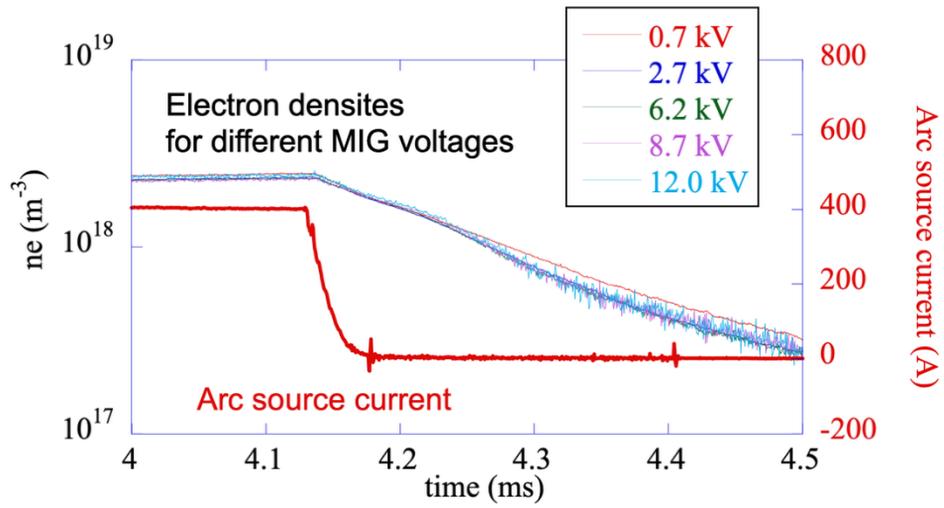

**Fig. 3** Line averaged electron density variation in time before and after termination of arc plasma injector at t =4.13 ms, for five different MIG voltages from 0.7 kV to 12.0 kV with 2 kG of cusp B-field strength. The thick red line shows the arc source current for plasma injection, and five thin colored lines show the electron density variation at different MIG voltages. Note that the MIG voltages were maintained at constant values until t=6.2 ms

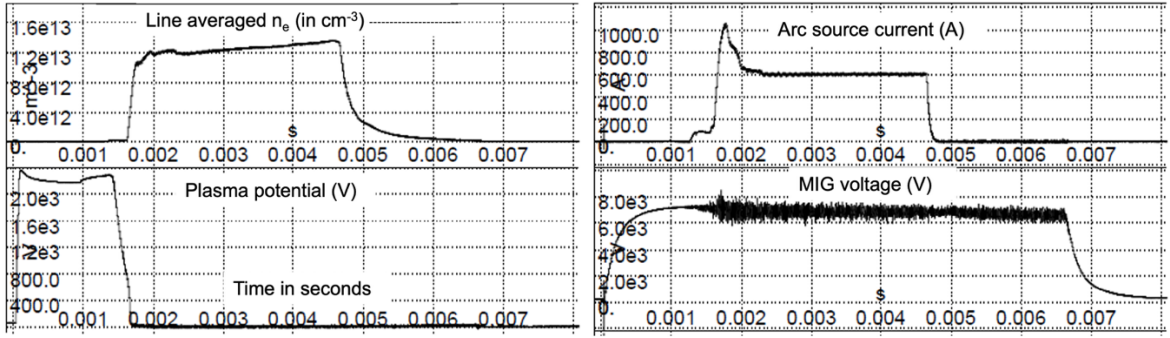

**Fig. 4** Plasma potential variation (bottom left) measured by an emissive probe at the center of the cusp coil during the WB-8 shot. The experimental setup and cusp plasma condition are shown for the plasma injection by the arc source (top left) and MIG voltage (bottom right) with 2 kG cusp B-field, and the line-averaged electron density (top right, in units of $cm^{-3}$)

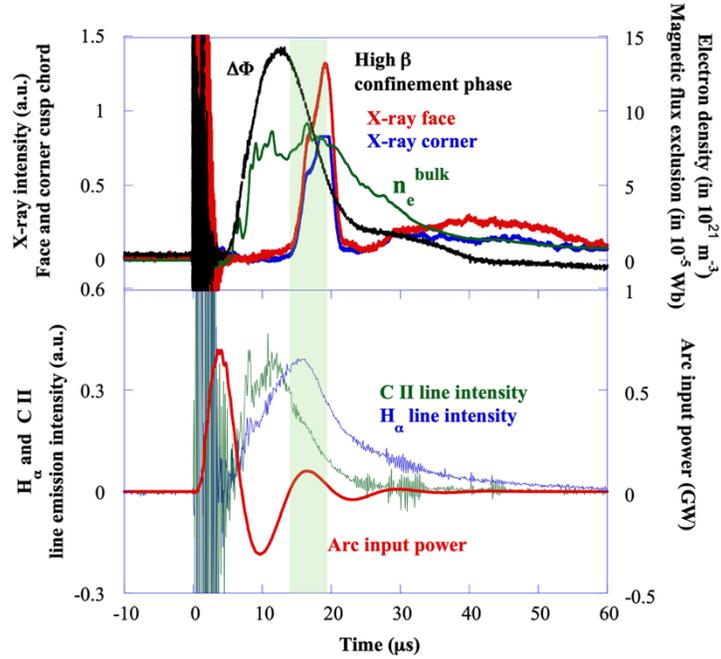

**Fig. 5** Time-resolved experimental results showing enhanced high-energy electron confinement in the high beta plasma state in WB-X, as highlighted by the light green region between 14 μs and 19 μs. The top figure shows the hard x-ray emission above 2 keV intensities from two lines of sight through the central region of the cusp, the cusp plasma density, and the magnetic flux exclusion resulting from plasma diamagnetism. The bottom figure shows the input power for the coaxial arc plasma injectors as well as the temporal variation of carbon and hydrogen emission intensities. Note that the duration of the high beta state in WB-X was limited to ~5 μs, due to the high impurity content from the use of polypropylene film and less-than-ideal crowbar operation of the pulse power system, as shown by the ringing nature of arc plasma power coupling. Reproduced from Ref. 12

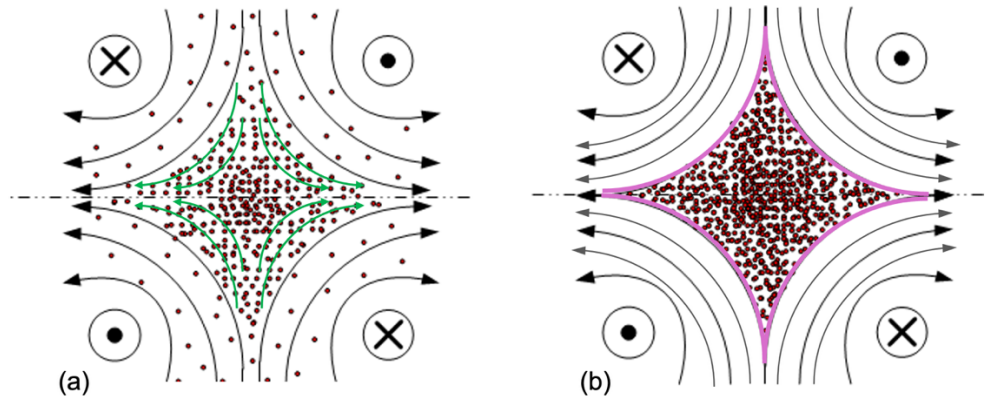

**Fig. 6** Comparison of magnetic cusp systems operating at (a) low beta plasma condition and (b) high beta plasma condition. At a low beta, the plasma pressure is insufficient to exclude the magnetic field generated by the coils within the plasma volume. Some magnetic field lines are colored green for clarity. In comparison, the high beta plasma can exclude the magnetic field from the coils in the plasma volume due to the strong diamagnetic current layers (shown as thick purple lines) at the boundary between the field-free core region and the nearly plasma-free outer region

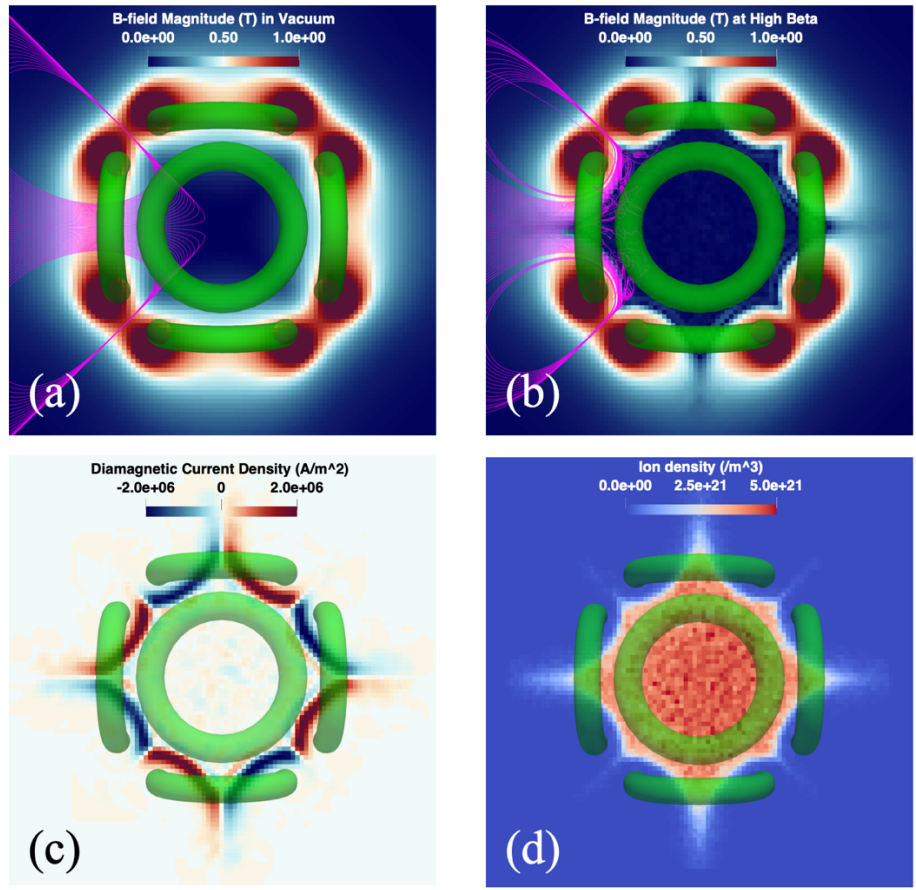

**Fig. 7** 3D Particle-in-Cell simulation results showing the evolution of magnetic field topology in the 6-coil magnetic cusp system after dense plasma injection, leading to boundary layer diamagnetic currents and high beta cusp plasma confinement with a steep density gradient at the cusp boundary. The plots are generated on the 2D plane across the middle of the 6-coil cusp system. (a) Contour plot of vacuum magnetic field magnitude with magnetic field lines connecting inner and outer regions of the cusp system. (b) Contour plot of magnetic field magnitude at a high beta state showing exclusion of the magnetic field in the inner region of the cusp. (c) Diamagnetic current layers separating two regions of the high beta cusp plasma, with the blue color for the current vector into the plane and red color for out of the plane. (d) Plasma density profile with sharp density gradients at the current layer locations, showing confinement property of the high-beta cusp plasma

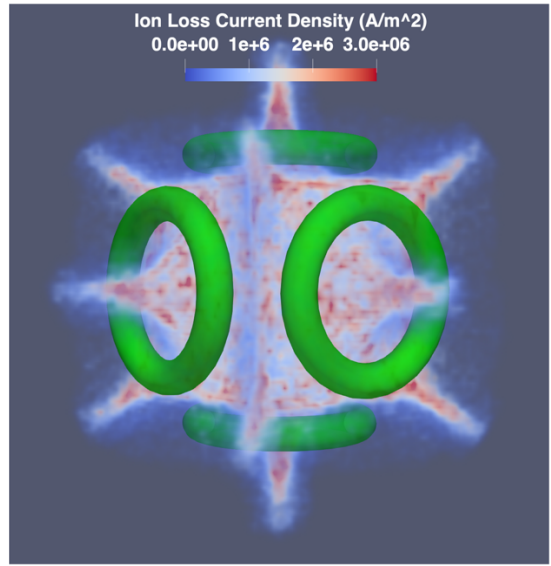

**Fig. 8** Spatial profile of cusp plasma loss rates (i.e., the amplitude of ion current density escaping out of the cusp system) at a high beta state in a 3D volume plot, showing localized losses around the centers of the coils and the corners of the cube

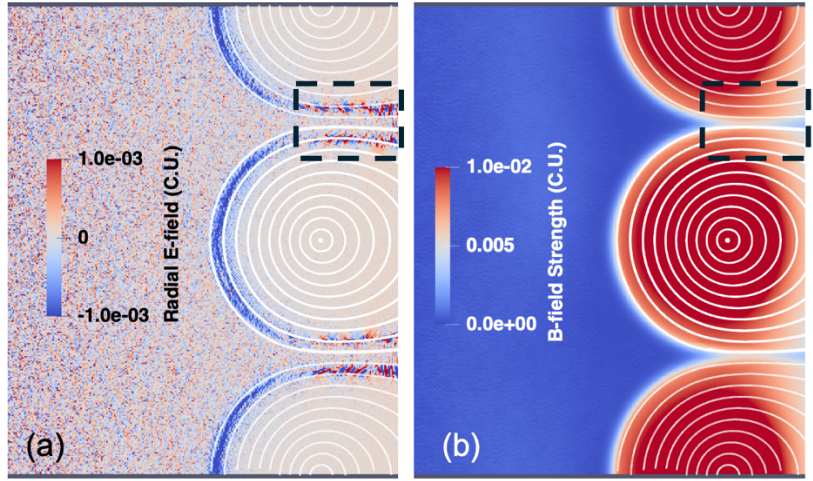

**Fig. 9** Location of potential plasma instability occurrence based on radial electric field profile from a high-beta picket fence simulation, as highlighted by the dotted boxes. (a) Radial electric field profile and (b) magnetic field strength profile. Note that thick white lines are the magnetic field lines of the high beta state, and the use of dimensionless code units in the plots, as described in the previous study [26]

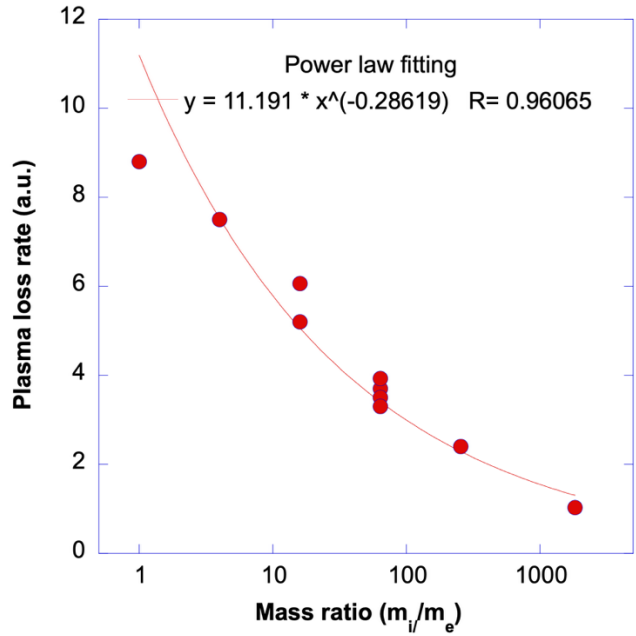

**Fig. 10** Plasma loss rate of high beta picket fence as a function of mass ratio between ions and electrons from the PIC simulations of the previous study [26]

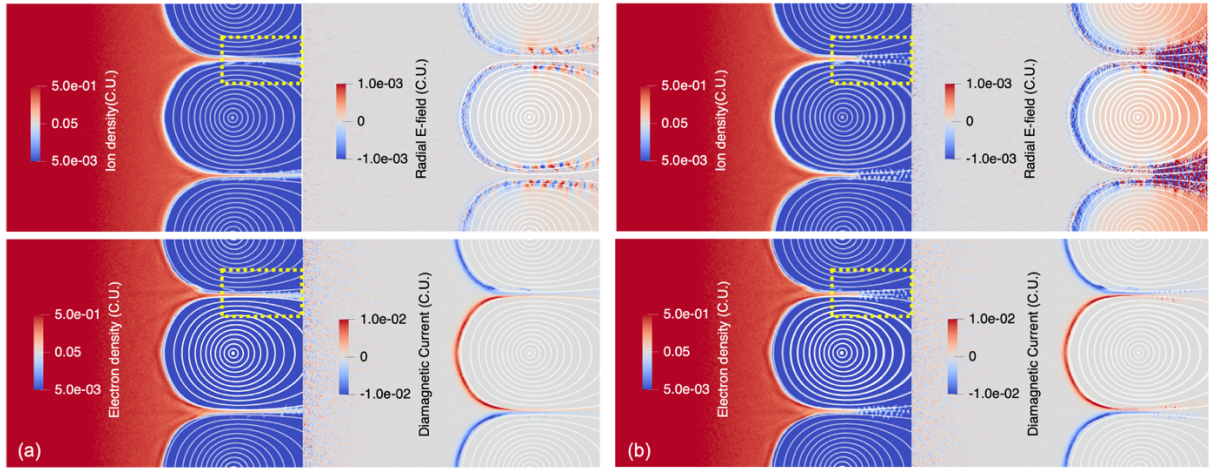

**Fig. 11** Improvement in plasma confinement with the injection of electron beams to the high beta picket fence, as shown in the change in plasma density profile. (a) Plasma properties of a high beta picket fence without electron injection, showing density profiles of ions (top left) and electrons (bottom left), boundary layer radial electric fields (top right), and diamagnetic current layer (bottom right). (b) Plasma properties of a high beta picket fence with electron injection, showing density profiles of ions (top left) and electrons (bottom left), boundary layer radial electric fields (top right), and diamagnetic current layer (bottom right). Note the significant density reduction by about an order of magnitude in the outer region highlighted by the yellow dotted boxes for the case with electron beam injection (b), compared to the case without electron beam injection (a)